# Investigation of Cathode Electrolyte Interphase Formation via Coupling Liquid Electrochemical TEM and GC/MS

Kevyn Gallegos-Moncayo[1], Josephine Rezkallah[1,4], Justine Jean[1,5],
Arash Jamali[1,4], Grégory Gachot[1,2*], Arnaud Demortière[1,2,3*]

[1]Laboratoire de Réactivité et de Chimie des Solides (LRCS), CNRS UMR 7314, UPJV, Hub de l'Energie, Rue Baudelocque, Amiens, France.
[2]Réseau sur le Stockage Electrochimique de l'Energie (RS2E), CNRS FR 3459, Hub de l'Energie, Rue Baudelocque, 80039 Amiens Cedex, France.
[3]ALISTORE-European Research Institute, CNRS FR 3104, Hub de l'Energie, Rue Baudelocque, 80039 Amiens Cedex, France.
[4]Plateforme de Microscopie Electronique, UPJV, Hub de L'énergie, 15 rue Baudelocque, 80039 Amiens, France

Corresponding Author: *arnaud.demortiere@cnrs.fr, *gregory.gachot@u-picardie.fr

**ABSTRACT**

A deeper understanding of the cathode electrolyte interphase (CEI) formation mechanism is essential to elucidate battery degradation. Here, we combine Liquid Electrochemical Transmission Electron Microscopy (ec-TEM) with Gas Chromatography/Mass Spectrometry (GC/MS) to monitor CEI evolution in a realistic electrochemical environment, focusing on electrolyte behavior under high voltages. The correlation between the electrochemical response, gas and liquid analysis after cycling, and the observation of deposited species on the working electrode (WE) reveals the processes governing CEI formation, stability, and composition. Cycling between 4 and 6 V vs Li leads to dispersed particles (1-1.5 μm) instead of a continuous film. These are partly composed of LiF and an amorphous phase that prevents dissolution at high potential. When cycled between 2.5 and 5.5 V, an anodic current peak indicates the formation of a 36 nm amorphous thin film without crystalline LiF, attributed to EC oxidation producing HF and subsequent LiF at a higher potential. LiF dissolution appears to follow a two-step pathway: electrolyte oxidation forms soluble intermediates, which are later reduced at lower potential to yield species capable of dissolving LiF. These results provide new insights into CEI formation and dissolution mechanisms, underscoring the need for further studies across different potential windows and with non carbonate electrolytes to validate these findings.

# INTRODUCTION

Cathode Electrolyte Interphase (CEI) is a thin layer, on the order of tens of nanometers, formed as a result of the oxidation of the liquid electrolyte at the positive electrode in batteries[1,2,3]. The oxidation of the electrolyte in the cathode side takes place when the lowest unoccupied molecular orbital (LUMO) energy level of the cathode is lower than the highest occupied molecular orbital (HOMO) energy level of the electrolyte[4,5]. Transition metals (TM) tend to dissolve into the electrolyte, reducing the efficiency and lifespan of the cathode itself, and the deposition on the anode can lead to the formation of metallic spots that become nucleation points for Li-dendrite growth[6,7]. Moreover, CEI rupture or dissolution leads to its reformation once contact between electrolyte and cathode is reestablished, affecting the electrolyte and reducing the lifespan and quality of the battery[8,9,10].

The formation and composition of the CEI have not been as widely studied as those of the Solid Electrolyte Interphase (SEI). In recent years, growing attention has been devoted to the CEI due to its critical role[11,12], and it has been investigated using various techniques such as XPS, SEM, NMR, and TEM[13,14,15,16]. Among these methods, TEM-based techniques offer a major advantage, enabling direct and precise observation of the CEI layer, as well as detailed compositional analysis at different levels, providing comprehensive insights into its multiple components. To reproduce conditions as close as possible to those of a real battery system, liquid electrochemical TEM (ec-TEM) combined with a dedicated electrochemical sample holder has been employed to investigate not only CEI formation[17], but also other key phenomena in batteries, such as SEI formation, parasitic phase generation, and active material evolution[18,19,20,21,22].

These studies are made possible by assembling a microbattery within the sample holder, which allows electrochemical measurements to be performed directly inside the microscope. In the work of Bhatia *et al.*[23] FIB-lamella of LNMO was used as a cathode in a microbattery. Cracks and grain size reduction due to amorphization of LNMO were observed, revealing key factors contributing to the rapid capacity fade of the battery. Regarding the electrolyte/electrode interaction, Dachraoui *et al.*[24] reported that the nucleation and evolution dynamics of the SEI proceed through a two-step mechanism: first, the formation of inorganic nanoparticles, followed by their growth and transformation into a mosaic-type structure composed of both organic and inorganic components, where the inorganic phases include $LiF$, $Li_2O$, $LiOH$, and $Li_2CO_3$.

A comprehensive understanding of the Cathode Electrolyte Interphase (CEI) and its role is essential to advance high energy density batteries operating at high potentials, such as those based on layered materials (LMO, LNMO, NMC, LCO)[25,26]. The formation and stability of the CEI are influenced by multiple factors, including the composition of the cathode or anode[15,27], the electrolyte formulation[28,29], and the applied cut-off voltages[30].

In this study, we employ a coupled liquid-electrochemical TEM and GC/MS system, using a dedicated sample holder and microchip designed for microbattery assembly, to investigate the formation of the cathode/electrolyte interphase (CEI). This correlative approach, previously demonstrated on Ni-based systems for studying SEI formation, has proven effective in identifying oxidation/reduction products and interfacial layer growth[31]. In addition, STEM-EDX and 4D-STEM ACOM analyses were conducted to further elucidate the crystalline phases present within the CEI layer[32].

## RESULTS AND DISCUSSION:

The first study carried out was the electrochemical LP30 cycling using a glassy carbon(GC) and working electrode (WE). Figure 1a shows a schematic representation of the top chip used for microbattery (MB1). After cycling and subsequent cleaning, the cathode was examined in STEM mode (Figure 1b). The images reveal the presence of a non-continuous layer around the electrode, along with additional surface deposits. These observations suggest that electrolyte oxidation induces the formation of a cathode–electrolyte interphase (CEI) layer through the deposition of insoluble decomposition products on the electrode surface.

The electrochemical curve obtained in this study (Figure 1c) closely resembles that reported by Azcarate *et al.*[33], where the formation of a surface layer composed of both inorganic and organic products was attributed to the oxidation of DMC and EC components of the LP30 electrolyte, along with the presence of LiF. In our case, the electrochemical curve exhibits an oxidative behavior beginning at approximately 5 V vs Li+, whereas in Azcarate's work, the onset of LP30 oxidation occurred at 4.5 V vs $Li^+$. In our case, a system with a pseudo-reference in Pt was used, which explains the variation in voltage between studies. To have a better understanding of the composition of the layer found in the GC cathode of MB1, STEM-EDX (cf. Figure 1d-f) was performed.

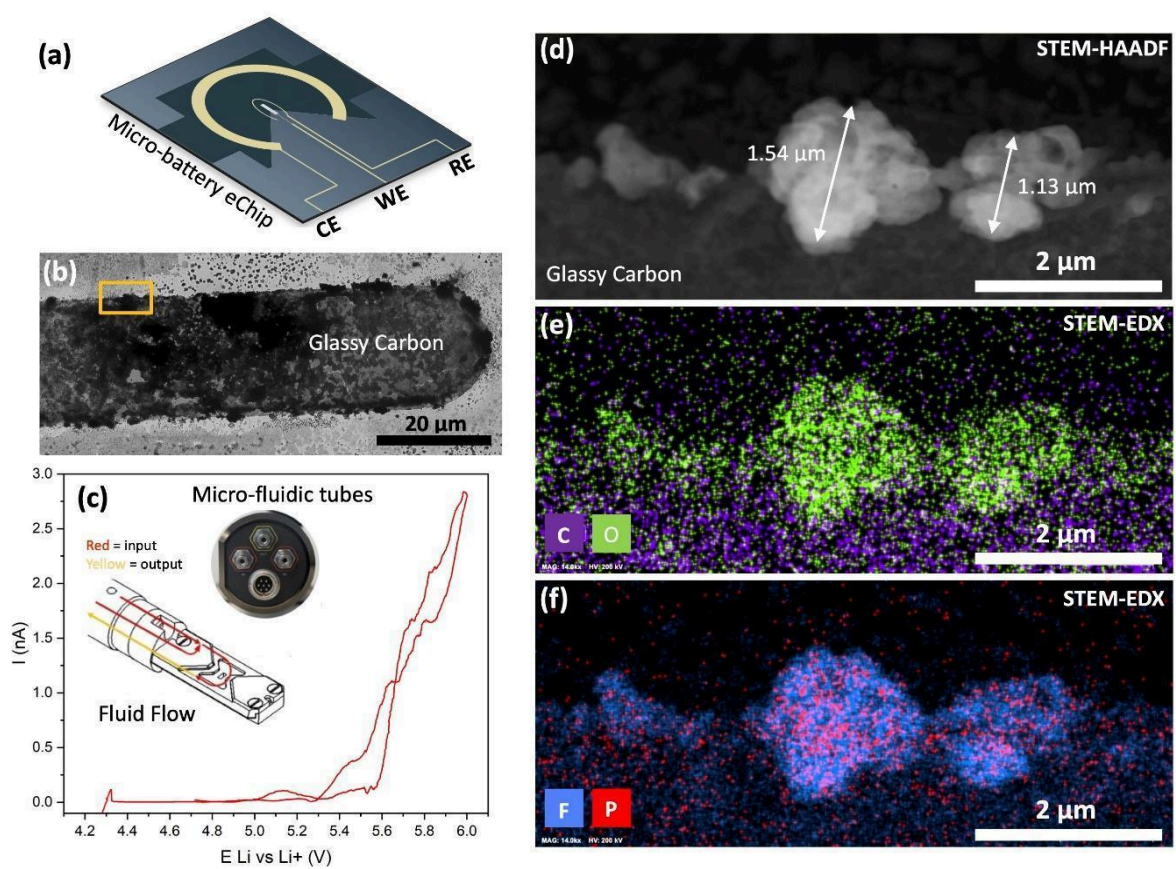

*Figure 1:* (a) schematic representation of the sample holder head, with WE, CE, and RE for electrochemistry test and observation in TEM from Protochips, (b) general view of the GC electrode remarking the zone in orange where the analysis were carried out, (c) electrochemistry curve obtained after the first cycle of LP30 inside the micro battery system, smoothed via Origin software with a smoothing value of 100, and the schematic representation of the micro-fluidic tubes, presenting in red the double input and in yellow the output, (d) STEM-HAADF image of a selected particle for STEM-EDX analysis, (e) STEM-EDX elemental map of particles highlighting C and O, the particles are mainly formed by O, (f) STEM-EDX elemental map of particles highlighting F and P, the particles are mainly formed by F.

The STEM-EDX data reveal a discontinuous layer formed by particles ranging from 1 to 1.5 µm. The predominant components are O and F. Quantitative analysis (Table S1) indicates that fluorine is the predominant element, representing approximately 73 at.%, while oxygen accounts for about 15 at.%. The detection of around 2 at.% of phosphorus confirms that the observed particles do not correspond to residual $LiPF_6$ from the electrolyte, but rather to newly formed compounds. These findings are consistent with the formation of a LiF-rich CEI layer, as commonly reported in the literature.

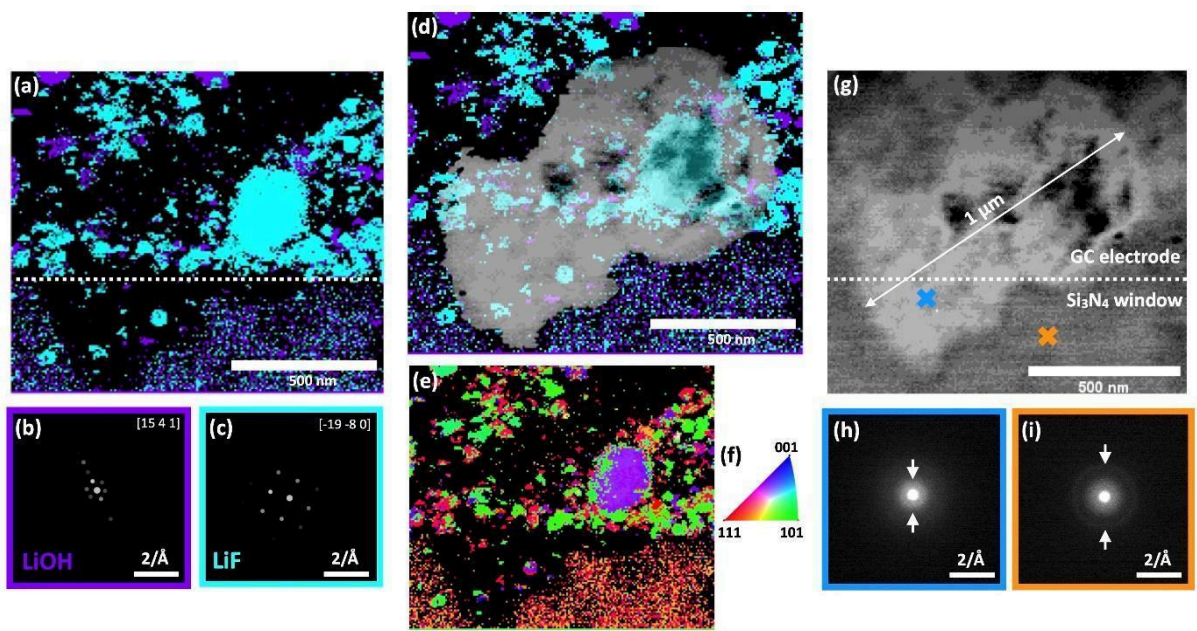

*Figure 2:* Analysis of an edge particle in the GC electrode with (a) 4D-STEM ACOM phase mapping of the particle at the edge of the GC (white dotted line) with both LiOH attributed phase (purple) and LiF attributed phase (cyan), (b) DP after e-Pattern cleaning of LiOH, (c) DP after e-Pattern cleaning of LiF, (d) an overlay of the amorphous and crystalline particle with the 4D-STEM phase map to present the complete particle morphology and the spatial distribution of the amorphous and crystalline phases over it, (e) 4D-STEM orientation mapping of the particle at the edge of the GC, (f) 4D-STEM orientation color map, (g) the particle presenting the amorphous zone (gray) and crystalline zone (black) distinction, (h) DP of the particle amorphous zone (no cleaned), (i) DP of the exterior amorphous zone identified as silicon nitride window of the top chip of the micro battery (no cleaned).

In the 4D-STEM ACOM (Automated Crystal Orientation Mapping) analysis (Figure 2), a particle approximately 1 µm in size was examined. Both crystalline and amorphous regions were identified at the electrode edges. For the amorphous areas, two distinct diffuse ring patterns were observed, one corresponding to the particle itself and another to its surroundings (Figure 2g). The external diffuse ring can be attributed to the $Si_3N_4$ chip window (Figure 2i), whereas the internal, broader, and more diffuse ring corresponds to the particle interior (Figure 2h), suggesting the presence of an organic amorphous phase likely associated with the CEI formation.

In the case of the crystalline phase, 4D-STEM analysis revealed that the particle is partially crystalline (Figure 2a) and two possible components of the CEI were identified: LiF (Figure 2c) and LiOH (Figure 2b). The presence of LiF, commonly reported in the literature, arises from the degradation of the electrolyte salt[34]. Figure 2d shows an overlay of the amorphous and crystalline regions, revealing a strong correspondence within the central black area. This agreement is further confirmed by the orientation maps of the particle (Figures 2e–f), which highlight the consistency in the crystalline zone.

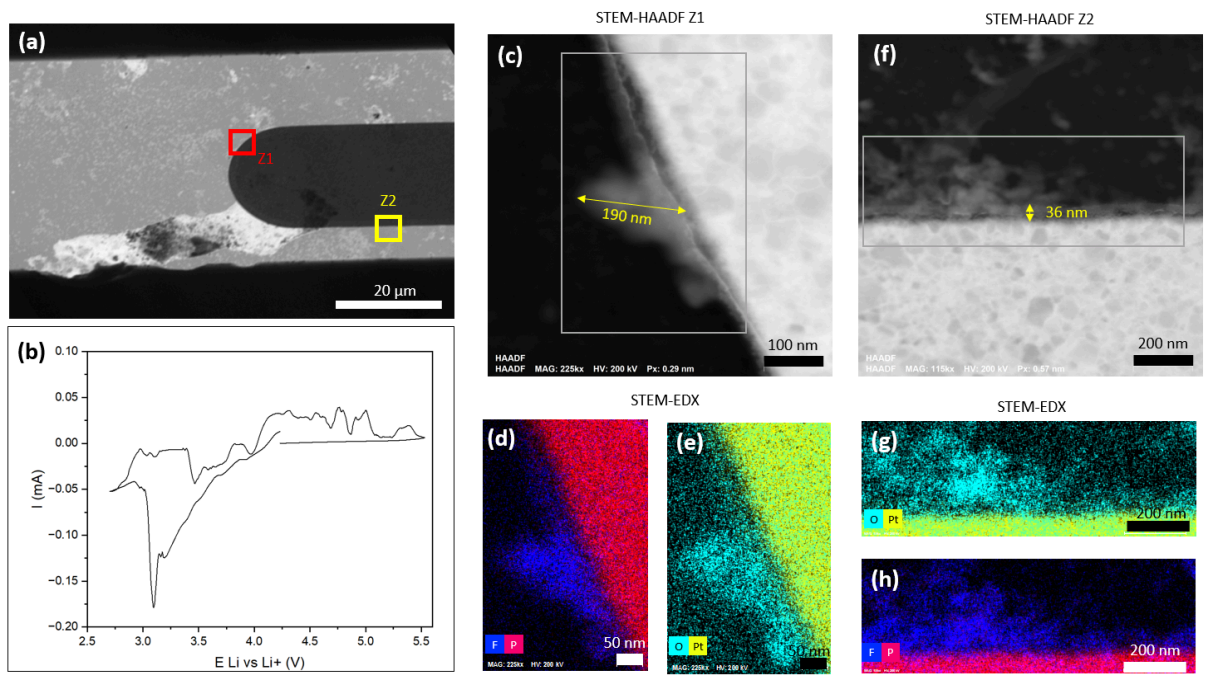

*Figure 3:* STEM-EDX analysis of the CEI layer obtained over Pt WE, (a) general view of the Pt WE and the two zones (Z1 in red and Z2 in yellow) where analysis were carried out, (b) electrochemical curve obtained in first cycle of LP30, smoothed in origin software with a smoothing value of 100, (c) Z1 STEM-HAADF image with the gray square featuring the particle where quantification and STEM-EDX analysis was carried out, (d) STEM-EDX elemental map presenting F and P with the particle being formed mostly by F, (e) STEM-EDX elemental map presenting O and Pt with the particle being formed mostly by O, (f) Z2 STEM-HAADF image with the gray square featuring the zone where quantification and STEM-EDX analysis was carried out and presenting the CEI layer, (g) STEM-EDX elemental map presenting O and Pt with the CEI layer being formed mostly by O, (h) STEM-EDX elemental map presenting F and P with the CEI layer being formed mostly by F. It is important to remark that P is present in the elemental map due to its proximity with Pt in energy.

For the second part of the study, a Pt working electrode was employed, and the cycling window was shifted toward lower voltages. Figure 3a shows a low-magnification view of the electrode, highlighting two distinct regions selected for analysis. In the corresponding electrochemical curve (Figure 3b), several peaks are observed. Despite some instability in the measurement, two dominant peaks can be clearly identified, corresponding to the oxidation processes of the liquid electrolyte.

The first anodic peak, appearing at approximately 4.77 V vs Li, is attributed to the oxidation of EC, while the second peak, near 5.0 V vs Li, corresponds to the oxidation of DMC. A third, more distant peak at about 5.37 V vs Li is hypothesized to arise from the oxidation of soluble organic intermediates derived from DMC[33]. A cathodic peak of −0.18 mA was also observed, likely associated with a reduction process. After this initial reduction event, no additional peaks were detected in the next cycles. This behavior contrasts with that reported in the previous study, where the potential window did not extend to such low voltages. Post-cycling analysis of the chip revealed no detectable crystalline phase.

The STEM-HAADF analyses of zone 1 (Z1) and zone2 (Z2) reveal a thin layer, suggesting the formation of a CEI layer. The CEI layer appears as a continuous inorganic layer with a thickness of 36 nm, with some particles deposited over its surface (Figure 3c and 3f). The layer is primarily composed of F and O (Figure 3d and 3e). Although P signals were detected over the electrode, it is important to note that the EDX energies of Pt and P are very close, meaning the observed signal could originate from Pt rather than P. To verify this, quantitative analyses were performed for Z1 (Table S2) and Z2 (Table S3). In both cases, no P or C was detected, confirming that the layer mainly consists of amorphous F- and O-based species.

In the study by Zhang *et al.*[17], a similar investigation was performed using Ti electrodes and an electrolyte based on a different carbonate solvent, *i.e.* propylene carbonate (PC). They observed the formation of a nanometric LiF layer during charging. According to Zhang's group, this LiF originates from the oxidation of PC in the presence of $LiPF_6$. Their DFT calculation indicated that the oxidation of the carbonate promotes the degradation of $PF_6^-$ leading to HF generation and further LiF formation.

In our case, on the GC electrode, where the potential never dropped below 4 V, no continuous thin layer was observed. Instead, a discontinuous accumulation of both amorphous and crystalline particles formed. Because the electrode never reached a sufficiently low potential for LiF dissolution (≈ 3 V vs Li), LiF remained stable and coexisted with oxidation products from EC/DMC, leading to the formation of a mixed amorphous–crystalline CEI. The LP30 electrolyte used during our experiment contains DMC/EC carbonates, and the oxidation mechanism were evaluated to determine whether LiF formation arises from their oxidation in the presence of $LiFP_6$.

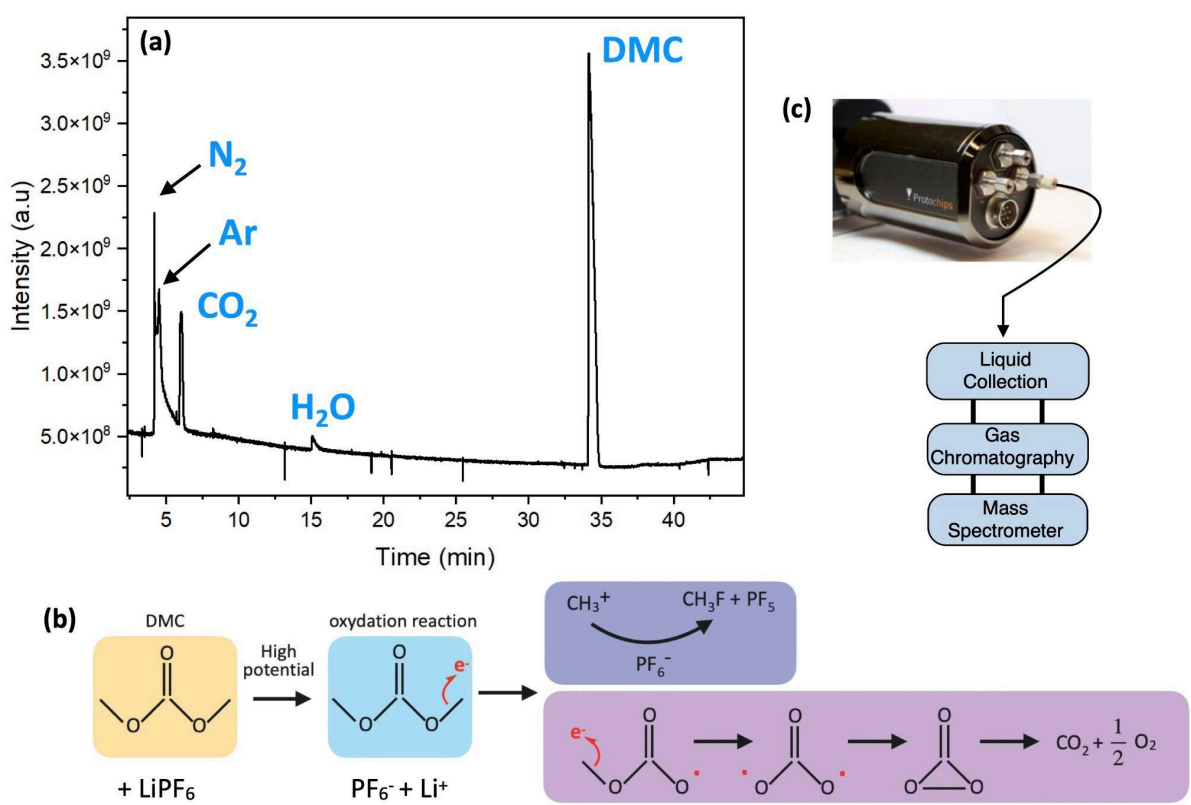

*Figure 4:* (a) gas analysis of GC/MS with the proposed mechanism for the oxidation of DMC. Presence of $N_2$ at 4.21 min, Ar at 4.52 min and $H_2O$ at 15.08 min, these gases are naturally present in air and are trapped in the needle of the injection syringe during injection, in the case of $CO_2$ at 6.06 min, the quantity is higher from what is expected of an air pollution, which confirms the presence of this gas due to oxidation of LP30, (b) mechanism proposition for the oxidation of DM and its eventual oxidation to $CO_2$ and $O_2$, as well as the indirect production of $CH_3F$ and $PF_5$ as a result of the e- liberation, (c) real image of the sample holder output and the schematic representation of the liquid collection and GC/MS analysis.

The DMC oxidation mechanism is illustrated in Figure 4a. In step 1, the DMC molecule undergoes oxidation, releasing one electron and a $CH_3^+$ ion, which is compensated by $PF_6^-$. The interaction between these species leads to the formation of two stable products: $PF_5$ and $CH_3F$. In step 2, a subsequent oxidation of DMC occurs, releasing another electron and generating a double radical intermediate (step 3). This intermediate then reacts to form a bond between the two radical oxygen atoms (step 4). However, the resulting molecule is unstable and decomposes into ½ $O_2$ and $CO_2$ (step 5). Remarkably, the mechanism proceeds without HF formation, with $CO_2$ clearly identified as the dominant oxidation product.

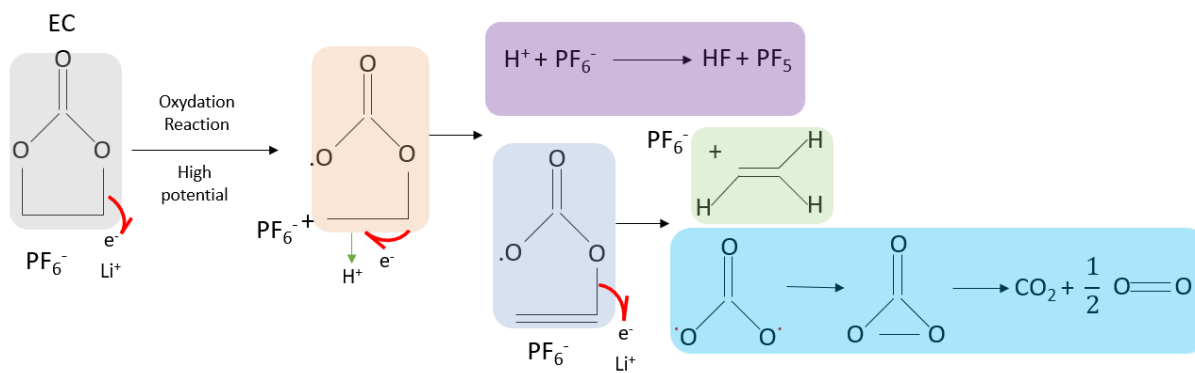

*Figure 5:* Schematic representation of the proposed mechanism of EC oxidation and eventual HF and LiF formation during cycling in LP30 electrolyte.

For the EC oxidation mechanism, a schematic representation is presented in Figure 5. Once the oxidation potential is reached, the EC molecule breaks, releasing an electron (stabilized by a Li+ ion) and liberating a $PF_6^-$. An $H^+$ is also released from the molecule, leaving a radical species behind. The $H^+$ stabilizes with a $PF_6^-$ and forms HF and $PF_5$. Since EC is a symmetric molecule, the radical intermediate can undergo a second oxidation, releasing another electron (again stabilized by a $Li^+$ ion and accompanied by $PF_6^-$), producing a positively charged molecule (stabilized by a $PF_6^-$) and a double radical molecule, similar to the mechanism observed for DMC. The overall process ultimately leads to the liberation of $CO_2$ and $O_2$.

The gas and liquid collected after cycling were analyzed by Gas Chromatography/Mass Spectrometry (GC/MS) (Figure 4c). The liquid analysis revealed no typical electrolyte reduction products (Figure S1), while the gas analysis (Figure 4a) exhibited a distinct $CO_2$ peak at 6.06 min, consistent with the proposed oxidation mechanism (Figure 4b) and confirming $CO_2$ as the primary oxidation product. Moreover, the probability of HF formation within the microbattery is low, as no $SiM_3F$ species, expected from a reaction between HF and the Si column, were detected in the chromatogram.

The study of Zhang *et al.*[17] also showed that during discharge, when Ti electrodes reached a potential of ≈3V vs Li, the formed LiF crystals disappeared. In our study, in the Pt electrode, where the voltage window was reduced under 3V, a corresponding reduction peak was observed. This finding is consistent with the previous observation, as no LiF particles were detected.

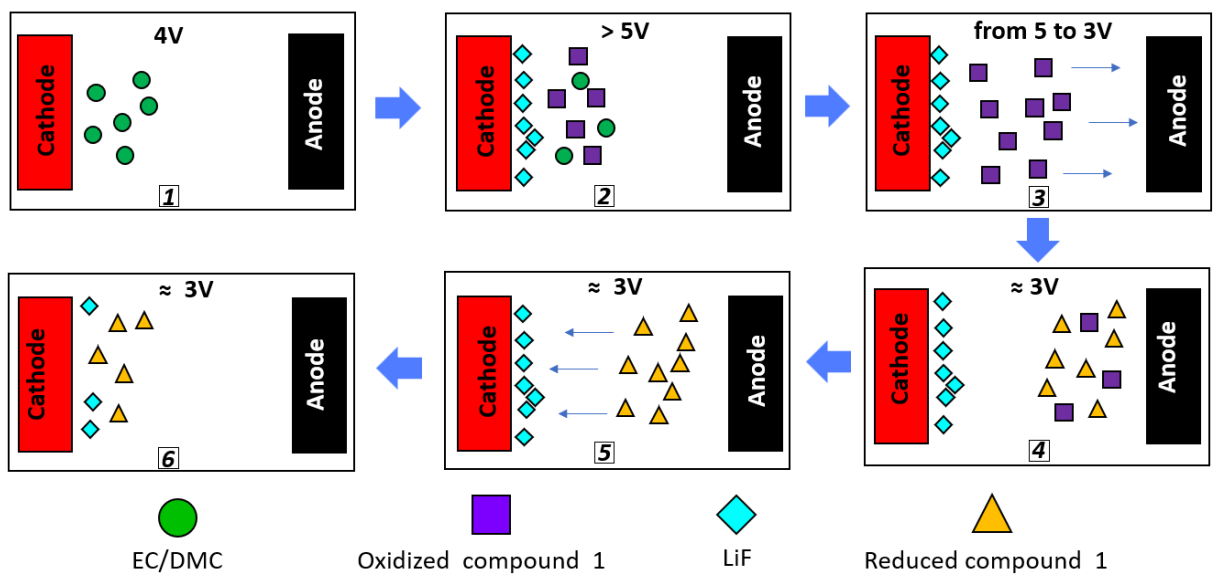

*Figure 6:* Schema of the two steps for the observation of LiF Formation and dissolution in LP30. 1.-OCV state of the electrolyte before cycling, step: 2.- oxidation reaction of DMC and EC, and LiF formation; 3.- migration of the soluble oxidation products to the anode. Step 2: 4.- reduction of the oxidized products, 5.- Migration of the soluble reduction products to the cathode, and 6.- dissolution of LiF.

The disappearance of LiF can be explained by its enhanced solubility in EC and reduced solubility in DMC, suggesting that solvent composition governs its stability[35]. An equilibrium is established in the electrolyte, preventing significant LiF dissolution. When LP30 enters an oxidative zone, unidentified oxidation products form, both soluble and insoluble products. The insoluble phase deposit near the cathode (CEI) and the soluble phase migrate toward the cathode. Upon decreasing the potential below a specific threshold, the oxidative products are reduced, as reflected by the anodic current, and subsequently migrate toward the cathode, where their reduced form favors the dissolution of LiF into the electrolyte.

To enable the formation and subsequent dissolution of LiF, a multi-step redox sequence must occur, as schematically depicted in Figure 6. This process involves (i) the oxidative decomposition of carbonate species, (ii) the generation of HF leading to LiF formation, (iii) the migration of oxidative by-products toward the anode, (iv) their electrochemical reduction, and finally (v) the reverse diffusion of the reduced species back to the cathode, where they modify the LiF dissolution/precipitation equilibrium. Such a dynamic exchange requires sufficiently high potentials to drive both oxidative and reductive reactions. In the present conditions (≈ 3 V), the applied potential remains below the threshold necessary to trigger these coupled interfacial processes. In contrast, in cells cycled within an extended voltage window (2.5 - 4.5 V), the formation and persistence of LiF have been unambiguously evidenced in previous studies[32], highlighting the strong voltage dependence of these mechanisms.

## Conclusion

The cathode/electrolyte interphase (CEI) plays a pivotal role in the long-term stability and safety of lithium-ion batteries by mitigating parasitic reactions and stabilizing the electrode/electrolyte interface. However, its formation and evolution mechanisms remain less understood compared with the extensively studied solid electrolyte interphase (SEI) on anodes.

In this work, we successfully coupled liquid electrochemical transmission electron microscopy (ec-TEM) with gas chromatography mass spectrometry (GC/MS) to investigate, under realistic liquid conditions, the mechanisms driving CEI formation and dissolution in LP30 electrolyte. Complementary 4D-STEM ACOM, STEM-HAADF, and STEM-EDX analyses provided nanoscale insights into the morphology, structure, and composition of the interphase.

Our results demonstrate that, within the typical voltage range of 3–5 V vs Li, LiF forms during charging and partially dissolves upon discharge, highlighting a reversible interphase behavior governed by redox-dependent solubility equilibria. The presence of amorphous and crystalline phases (LiF, LiOH, and organic residues) was confirmed by 4D-STEM ACOM mapping. In addition, GC/MS analysis and electrochemical evidence support the oxidation of carbonate solvents, particularly ethylene carbonate (EC), as the source of HF and, consequently, LiF formation through $LiPF_6$ degradation.

Although HF formation was not directly detected by GC/MS, likely due to quantities below detection limits or rapid reaction with oxidation/reduction products, the combined analytical evidence supports its transient generation during cycling. The coexistence and reversibility of LiF at the CEI are therefore proposed to arise from a two-step mechanism involving: (i) oxidation of carbonates leading to LiF formation and (ii) subsequent reduction of soluble oxidative products capable of dissolving LiF at lower potentials.

These findings underline the dynamic and chemically heterogeneous nature of the CEI, controlled by the delicate balance between electrolyte oxidation and reduction pathways. Understanding the interplay between electrolyte composition, potential window, and CEI chemistry is key to improving high-voltage cathode stability.

Future work will focus on (i) performing analogous experiments using non-carbonate-based electrolytes to validate the proposed HF-driven LiF formation mechanism and (ii) cycling without the oxidation stage to isolate reduction-driven dissolution processes. Establishing such mechanistic clarity will guide the rational design of new electrolytes and additives to tailor CEI composition and enhance the durability and performance of next-generation high-voltage Li-ion batteries.

## Material and methods

### Micro battery and cycling conditions

Micro batteries are composed of a top and a bottom chip (Protochips, Morrisville, NC 27560 USA), both with an $N_3Si_4$ window separated by 150nm, and 3 electrodes, WE, RE, and CE. WE for MB1 is made of GC (ECT-45CR), and for MB2 WE is made of Pt (ECT-45PT). CE and RE are both made of Pt. The bottom chip is the same in both microbatteries (EPB-55DNF). The sample holder used was an ec-TEM liquid sample holder (Protochips, Morrisville, NC 27560 USA).

MB1 was cycled using a 3-electrode configuration, from 3 to 6 V vs Li, with a sipping rate of 1mV/s and LP30 flow of 1μl/min. MB2 was cycled with a 2-electrode configuration between 2.5 and 5.5 V vs Li with a sipping rate of 3mV/s and LP30 flow of 1 μl/min. Common LP30 was used for the batteries: lithium hexafluorophosphate (LiPF6) in a binary solvent system of ethylene carbonate (EC) and dimethyl carbonate (DMC) in a 1:1 volumetric ratio.

Micro batteries were cleaned while closed using DMC to cut any possible reaction with air. Then opened, and the top chip was dried in a vacuum and reassembled using a no-window bottom chip for easier observation in TEM.

### STEM-EDX analysis and TEM observation

TEM and STEM analysis were carried out using a Tecnai G2 F20 S-Twin (Thermo Fisher Scientific, Waltham, MA, USA) system using a OneView CMOS camera (Gatan, Pleasanton, CA, USA). The acceleration voltage was 200 kV. Images were acquired using an aperture C2 of 150 µm, and STEM-EDX analysis was carried out with an aperture C2 of 70 µm. Batteries were tilted at 20° for EDX data acquisition, with a spot size of 8. The acquisition of elemental maps was performed using an energy-dispersive X-ray spectroscope (EDX, Xflash, Bruker, Berlin, Germany). For quantification, the TEM Cliff-Lorimer method was used.

### 4D-STEM ACOM analysis

The 4D-STEM ACOM investigations were conducted using an accelerating voltage set to 200 kV. During the nano-diffraction experiments, the camera length was maintained at 300 mm. A precesion angle of 0.7° was used, aimed at minimizing dynamical scattering effects using NanoMegas device. The C2 aperture was precisely configured to 10 µm, resulting in a convergence semi-angle of 0.4 mrad. For electron beam control, Gun lens 3 was utilized, with the spot size adjusted to 5. The electron dosage for 4D-STEM analysis was established at 150 e/Å²/s. The resolution of each DP was configured to 512×512 pixels.

### GC/MS analysis

The acquisition parameters in GC/MS (Thermo Scientific): For GC, a flow rate of 1.5 mL.min$^{-1}$ of He and a temperature gradient[36]. For MS, an electron impact (EI) source with an ionization energy of 70 eV is used. Use of a quadrupole (Q) to separate compounds based on their mass/charge ratio (m/z) (identification of compounds being processed with the National Institute of Standards Library (NIST)).

### Software

Data processing of the 4D-STEM dataset was executed utilizing the ePattern[37] suite software, which facilitated denoising operations with a prominence value set at 5. Subsequently, the ASTAR software package (Nanomegas, Brussels, Belgium) was applied for the reconstruction of phase and orientation maps. This was achieved through the Automated Crystal Orientation Mapping (ACOM) technique, which relies on a pattern-matching algorithm.

**DATA AVAILABILITY:** For any data requirement, please contact the corresponding author.

### ACKNOWLEDGMENTS

As a part of the DESTINY PhD program, this publication is acknowledged by funding from the European Union's Horizon 2020 research and innovation program under the Marie Skłodowska-Curie Actions COFUND (Grant Agreement #945357). A part of the funding has been provided by the French Research Agency (ANR) as part of the DestiNa-ion Operando project (ANR-19-CE42-0014).

### AUTHOR CONTRIBUTIONS

A.D., K.G.M. have conceived the investigation. This work was supervised by A.D. The electrochemistry data and TEM data were acquired by K.G.M, J.J., A.J. and J.R. The 4D-STEM ACOM data was treated by K.G.M. and J.J. The GC/MS analysis was carried out by G.G. The manuscript was written by K.G.M. and A.D. All authors participated in the discussion and revision of this paper and finally approved this work.

**COMPETING INTERESTS**

The authors declare no competing financial or non-financial interests.